# Electronic structure of CaFe$_2$As$_2$: Contribution of itinerant Fe *3d*-states to the Fermi Level

E.Z. Kurmaev[1], J.A. McLeod[2], A. Buling[3], N.A. Skorikov[1], A. Moewes[2],

M. Neumann[3], M.A. Korotin[1], Yu.A. Izyumov[1], N. Ni[4], and P.C. Canfield[4]

[1]*Institute of Metal Physics, Russian Academy of Sciences-Ural Division,*
*620219 Yekaterinburg, Russia*

[2]*Department of Physics and Engineering Physics, University of Saskatchewan,*
*116 Science Place, Saskatoon, Saskatchewan S7N 5E2, Canada*

[3]*Department of Physics, University of Osnabrück, Barbarastr. 7,*
*D-49069 Osnabrück, Germany*

[4]*Ames Laboratory U.S. DOE and Department of Physics and Astronomy,*
*Iowa State University, Ames, Iowa 50011, USA*

We present density functional theory (DFT) calculations and a full set of X-ray spectra (resonant inelastic X-ray scattering and X-ray photoelectron spectra) measurements of single crystal CaFe$_2$As$_2$. The experimental valence band spectra are consistent with our DFT calculations. Both theory and experiment show that the Fe *3d*-states dominate the Fermi level and hybridize with Ca *3d*-states. The simple shape of X-ray photoelectron (XPS) Fe *2p*-core level spectrum (without any satellite structure typical for correlated systems) suggests itinerant character of the Fe *3d*-electrons. Based on the similarity of the calculated and experimental Fe *3d*-states distribution in LaOFeAs and CaFe$_2$As$_2$ we conclude that superconductivity in the FeAs-systems can be described within a minimal model, taking into account only Fe *3d*-bands close to the Fermi level.

# 1. Introduction

Soon after finding superconductivity in the $R$OFeAs ($R$ – rare earth) compounds [1] another family of Fe-As superconductors, $AE$Fe$_2$As$_2$ ($AE$=Ca, Sr, Ba, Eu) with the ThCr$_2$Si$_2$-type tetragonal structure (space group $I4/mmm$), was discovered [2-3]. A crystallographic phase transformation from a high temperature tetragonal phase to a low temperature orthorhombic phase occurs in these compounds between 140-170 K [2-3]. Although these materials are not superconducting at ambient pressure, superconductivity can be realized by doping the $AE$- and Fe-sites [2, 4-6] as well as by applying high pressure [7-9]. In both cases superconductivity can only occur when the phase transformation from tetragonal to orthorhombic structure can be easily realized or when it is fully suppressed. A similar structural phase transition also appears in the $R$FeAsO compounds and is likewise suppressed by F-doping [1], which strongly links the superconductivity of doped iron arsenides to the proximity to this structural instability of $AE$Fe$_2$As$_2$ compounds and possibly to the spin-density wave ordering [10-12].

The calculated valence band electronic structures of BaFe$_2$As$_2$ [13] and LaOFeAs [14] are similar. The bands around the Fermi level for both compounds are mainly formed by Fe $3d$-states. The difference in the valence band of both compounds is only observed in the non-iron states. In the LaOFeAs system, As $4p$-states are hybridized with O $2p$-states and separated from the Fe $3d$-bands whereas in BaFe$_2$As$_2$ the Fe $3d$- and As $4p$-bands are hybridized. Similar results are found from other DFT calculations [6, 13, 15]. In this respect the study of the electronic structure of CaFe$_2$As$_2$ is of particular interest because Ca $3d$-states can interact with Fe $3d$-states and therefore modify the distribution of Fe $3d$-states near the Fermi level. We have studied both iron and calcium resonant and nonresonant $L_{2,3}$ X-ray emission spectra (RXES, NXES), which probe partial Fe $3d$ and Ca $3d$ densities of states (DOS). We compare them with XPS valence band measurements (which probe the total DOS) and our full potential linearly-augmented plane wave (FP-LAPW) electronic structure calculations of $CaFe_2As_2$.

## 2. Experimental and Calculation Details

Single crystals of $CaFe_2As_2$ were grown out of a Sn flux using conventional high-temperature solution growth techniques in experimental conditions described in [3, 19]. Elemental Ca, Fe, and As were added to Sn in the ratio of $[CaFe_2As_2]$:Sn=1:48 and placed in a 2 ml alumina crucible. A second catch crucible containing silica wool was placed on top of the growth crucible and sealed in a silica ampoule under approximately 1/3 atmosphere pressure of argon gas. The sealed ampoule was placed in programmable furnace and heated to 850° C and cooled over 36 h to 500° C. The high quality of the $CaFe_2As_2$ crystals is confirmed by an extensive characterization employing X-ray diffraction, neutron diffraction, thermodynamic and transport techniques [10, 19].

XPS measurements were obtained using a Perkin-Elmer PHI 5600 ci Multitechnique System with monochromatized Al $K\alpha$ radiation (full width at half-maximum (FWHM) of 0.3 eV). The energy resolution of the spherical capacitor analyzer was adjusted to approximately $\Delta E = 0.45$ eV. The pressure in the ultra-high vacuum chamber was in the $10^{-10}$ mbar range during the measurements. The $CaFe_2As_2$-crystal was cleaved in situ. The surface contamination was monitored O $1s$ and C $1s$ core level spectra before and after our measurements.

The resonant and non-resonant X-ray emission measurements of $CaFe_2As_2$ were performed at the soft X-ray fluorescence endstation of Beamline 8.0.1 at the Advanced Light Source in the Lawrence Berkeley National Laboratory [16]. The endstation uses a Rowland circle geometry X-ray spectrometer with spherical gratings and an area sensitive multichannel detector. We have measured the resonant and non-resonant Fe $L_{2,3}$ ($3d,4s \rightarrow 2p$ transition) and non-resonant Ca $L_{2,3}$ ($3d,4s \rightarrow 2p$ transition) X-ray emission spectroscopy (XES). The instrument resolution for Fe $L_{2,3}$ and Ca $L_{2,3}$ X-ray emission spectra was 0.8 eV for Fe, 0.4 eV for Ca. All spectra were normalized to the incident photon current using a clean gold mesh in front of the sample to measure the intensity fluctuations in the photon beam. The X-ray absorption spectroscopy (XAS) measurements for Fe were taken in total fluorescence yield (TEY) mode to minimize surface oxidation effects, with a resolution of 0.3 eV. The excitations for the resonantly excited XES measurements were determined from the XAS spectra; the chosen energies corresponded to the location of the $L_3$ and $L_2$ thresholds and as well as one energy well above resonance.

All band structure calculations were performed within the FP-LAPW method as implemented in the WIEN2k code [17]. For the exchange-correlation potential we used the Perdew-Burke-Ernzerhof gradient approximation variant (GGA) [18]. The Brillouin zone integrations were performed with a $11 \times 11 \times 11$ special $k$-point grid and $R_{MT}^{min}K_{max}$=7 (the product of the smallest of the atomic sphere radii $R_{MT}$ and the plane wave cutoff parameter $K_{max}$) was used for the expansion of the basis set. The experimentally determined lattice parameters of the high temperature phase of $CaFe_2As_2$ ($a$=3.912 Å, $c$=11.667 Å) [19] were used in our calculations. We chose the $z$-parameter of As in BaFe₂As₂ [13] as a starting approximation and performed structural relaxation calculations to minimize internal forces. The resulting coordinates of As were (0.0, 0.0, 0.35814). Atomic sphere radii of $R_{Ca}$=2.5, $R_{Fe}$=2.17 and $R_{As}$=1.92 a.u. were chosen in the FP-LAWP calculation. These were selected so that the spheres are nearly touching.

## 3. Results and Discussion

The electronic structure calculations of CaFe₂As₂ are shown in Fig. 1. The As *4s*-states are concentrated at the bottom of the valence band (~11.2 eV). The As *4p*-like band shows a two-peak structure (at ~5.2 and 3.3 eV) and the upper band is strongly mixed with Fe *3d* and Ca *3d*-states. The top of the valence band (0-1 eV) is formed by Fe *3d* and Ca *3d*-states with the main spectral weight centered at 0.48 eV. Fe *3d*-states dominate at the Fermi level. This situation is similar to for the Fermi level of LaOFeAs [14]. The Fe *3d*-states are hybridized with As *4p*-states located around 3-4 eV below the Fermi level A similar situation takes place in BaFe₂As₂ [13]. We performed FP-LAPW band structure calculations for BaFe₂As₂ using the manner outlined in Ref. [13]. A comparison of the DOS curves for CaFe₂As₂ and BaFe₂As₂ is shown in Fig. 2. Note that the position and shape of Fe *3d* DOS are nearly identical for both compounds and are not altered when Ca is replacing Ba. The Ca *3d*-states are found to be strongly mixed with Fe *3d*-states and As *4p*-states and give a noticeable contribution (~10%) to the density of states near the Fermi level. A similar result is found for Ba *5d*-states but the contribution is considerably smaller (~6%). From the band structure calculations of CaFe₂As₂ and the comparison with those for LaOFeAs we conclude that the FeAs-layer in both systems determines the main features of their energy bands. Most importantly the Fe $3d$-states provide in both cases the main contribution to the density of states near the Fermi level.

Despite the fact that the DFT-calculations for $CaFe_2As_2$ reveal no major differences when compared with other compounds of this type, the superconducting behavior with pressure is significantly different in $CaFe_2As_2$ than in similar compounds. According to neutron scattering measurements [11, 20], superconductivity emerges at a pressure of P > 0.35 GPa [7] which is believed to be associated with the emergence of a "collapsed" tetragonal phase, in which the cell volume changes with the basic symmetry of the crystal structure remaining the same. The existence of this phase in the range of pressures where superconductivity exists is confirmed by electric resistivity [7, 12] and magnetic susceptibility measurements [12]. Within the collapsed phase the magnetic order disappears and the local magnetic moments are quenched [11,20]. The suppression of the magnetic order and local moments by high pressure creates the conditions required for the appearance of superconductivity. DFT calculations of the total energy in a given cell volume change have shown to favor the tetragonal phase at low temperatures and simultaneously the disappearance of magnetic moments [11, 20].

Isostructural phase transitions under high pressure are well-known in Ce, which is a strongly correlated metal. However $CaFe_2As_2$ should be considered as a weakly correlated material, and therefore the DFT calculations carried out in [11, 20] are sufficient. It still remains pertinent to understand why only $CaFe_2As_2$ exhibits the collapsed phase among the set of $AEFe_2As_2$ compounds. In connection with this more detailed studies of structural and magnetic state of $AEFe_2As_2$ ($AE$=Ba, Sr, Eu) compounds at high pressure within superconducting state are necessary.

The XPS survey spectrum of single crystalline of $CaFe_2As_2$, cleaved in ultra-high vacuum (Fig. 3), demonstrates a very low level of oxygen and carbon contamination and therefore their negligible contribution to the valence band spectrum. XPS valence band and core level spectra of $CaFe_2As_2$ are shown in Fig. 4. The XPS Fe *2p* core level spectrum lacks the satellite structure typical for correlated systems (for instance for FeO [21]) and the Fe $2p_{3/2}$ peak is sharp and similar to metallic iron [22]. A similar simple shape of Fe *2p* XPS is observed for LaOFeAs [23]. Therefore we note that in these FeAs-based systems, the Fe *3d* electrons are not localized but rather itinerant in nature.

The XPS VB measurements of $CaFe_2As_2$ show 4 distinct subbands located at ~11.9, 5.1, 3.4 and 0.4 eV. The binding energies of these subbands are in excellent agreement with the calculated partial DOS (Fig. 1). We also note that the XPS VB of $CaFe_2As_2$ is in agreement with photoemission measurements of $LaFeAsO_{1-x}F_x$ (x = 0, 0.06) [23]. Neglecting small differences due to the contribution of O *2p*-states in $LaFeAsO_{1-x}F_x$, the overall picture of the XPS VB near the Fermi level is very similar for both compounds and mainly determined by Fe *3d*-states of FeAs layers.

Fig. 5 shows RIXS Fe $L_{2,3}$ spectra of $CaFe_2As_2$. The two main bands located around 705 and 718 eV correspond to Fe $L_3$ (*3d4s* $\rightarrow$ $2p_{3/2}$ transitions) and Fe $L_2$ (*3d4s* $\rightarrow$ $2p_{1/2}$ transitions) normal emission lines separated by the spin-orbital splitting of Fe *2p*-states. Resonantly excited Fe $L_3$ RXES and nonresonant Fe $L_{2,3}$ NXES of $CaFe_2As_2$ and LaOFeAs are compared in Fig. 6. They are found to be almost identical (in shape and in energy) and we conclude that the experimentally determined distribution of occupied Fe *3d*-states is similar in both compounds, as suggested by our electronic structure calculations. We have not found any essential changes in the full-width-at-half-maximum (FWHM) of the Fe $L_3$ XES though the width of the Fe $3d$-band of $AE$Fe$_2$As$_2$ is expected to be larger than that of LaOFeAs due to the shorter Fe-Fe distances [14].

The position of the Fermi level is determined by using the XPS Fe $2p$ binding energy for CaFe$_2$As$_2$ ($E_b$ = 706.8 eV). The intensity maximum of Fe $L_3$ XES is located within 1.25 eV of the Fermi level, demonstrating that the Fe $3d$-states dominate the top of the valence band for both one-layered and two-layered FeAs-systems. The Fe $L_3$ XES do not show any features that would indicate the presence of a lower Hubbard-band or a sharp quasi-particle feature that were predicted by the LDA+DMFT analysis [24].

The non-resonant Fe $L_{2,3}$ XES of LaOFeAs and CaFe$_2$As$_2$ presented in Fig. 7 show almost identical I($L_2$)/I($L_3$) intensity ratios. For free atoms the relative intensity ratio of $L_2$ and $L_3$ XES lines is determined only by the statistical population of $2p_{1/2}$ and $2p_{3/2}$ levels and therefore should be equal to ½. In metals the radiationless $L_2L_3M_{4,5}$ Coster-Kronig ($C$-$K$) transitions strongly reduce the I($L_2$)/I($L_3$) ratio [25] and therefore the I($L_2$)/I($L_3$) ratio is a measure for the metallicity of a transition metal compound. The intensity ratios shown in Fig. 7 for LaOFeAs and CaFe$_2$As$_2$ are more similar to those of metallic Fe than those of the strongly correlated FeO. This further supports our conclusion that the Fe $3d$-states in FeAs systems exhibit more itinerant than localized character.

Fig. 8 displays a comparison of the XPS VB, which probes the total DOS, and the Ca and Fe $L_3$ XES, which provides information about the partial $3d$-states of constituent atoms. The spectra are aligned using the Fermi level, which is determined for Fe and Ca $L_3$ XES using XPS Fe $2p$ and Ca $2p$ binding energies as displayed in Fig. 4. Fig. 8 shows that the experimental X-ray emission spectra are in good agreement with XPS VB and the calculated partial DOS. The most important aspect is that the Fe and Ca $L_3$ XES reside near the top of the valence band and therefore the electronic states in the vicinity of the Fermi level originate from Fe and Ca $3d$-states. Fig. 9b shows the theoretical XPS VB calculated by taking into account atomic photoionization cross-sections of the constituents. The calculation is in agreement with the experimental spectrum. This demonstrates that a DFT calculation is the proper approach to describe the electronic structure of $CaFe_2As_2$. In accordance with our electronic structure calculations (Fig. 2) the UPS of $BaFe_2As_2$ (E=2010 eV) [26], shown in Fig. 9a, is almost identical to the XPS of CaFe2As2.

To summarize, we have performed a full theoretical and experimental study of the electronic structure of $CaFe_2As_2$. Band structure calculations, experimental valence band spectra, and core level spectra show that Fe $3d$-states dominate at the Fermi level and are very similar to that of LaOFeAs. We conclude that the superconductivity of FeAs layered compounds may be described within the minimal model, taking into account only the essential Fe $3d$-bands that are close to the Fermi level [25].

## Acknowledgements

We acknowledge support of the Research Council of the President of the Russian Federation (Grants NSH-1929.2008.2 and NSH-1941.2008.2), the Russian Science Foundation for Basic Research (Project 08-02-00148), the Natural Sciences and Engineering Research Council of Canada (NSERC), and the Canada Research Chair program. PCC acknowledges useful discussions with S. L. Bud'ko and G. D. Samolyuk. Work at the Ames Laboratory was supported by the Department of Energy, Basic Energy Sciences under Contract No. DE-AC02-07CH11358.

**Figure Captions:**

Fig. 1. Calculated total and partial density of states for $CaFe_2As_2$.

Fig. 2. Comparison of $CaFe_2As_2$ and $BaFe_2As_2$ DOS calculations. In the bottom figure both the *4s*- and *4p*-states of As are plotted since they are energetically separated. In other cases the *4s*- and *6s*-states have minimal contribution and are omitted here.

Fig. 3. XPS survey spectrum of $CaFe_2As_2$.

Fig. 4. XPS VB and core level spectra of $CaFe_2As_2$. Arrows in the top figure (XPS VB) indicate distinct subbands at 11.9, 5.1, 3.4, and 0.4 eV.

Fig. 5. Fe *2p* XAS spectra (top panel) and $L_{2,3}$ RIXS spectra (lower panel) of $CaFe_2As_2$.

Fig. 6. Comparison of resonant Fe $L_3$ XES spectra of $CaFe_2As_2$ and LaOFeAs.

Fig. 7. Comparison of nonresonant Fe $L_3$ XES spectra of FeO, Fe-metal, $CaFe_2As_2$, and LaOFeAs. The ratio of the integral under the $L_2$ and $L_3$ peaks (the I($L_2$)/I($L_3$) ratio) is plotted in the upper right inset. Note how the ratios of $CaFe_2As_2$ and LaOFeAs are much closer to Fe metal then FeO.

Fig. 8. Comparison of XPS VB (a) with Fe $L_3$ RXES (b) and Ca $L_3$ XES (c) for $CaFe_2As_2$.

Fig. 9. Experimental XPS VB of $CaFe_2As_2$ and UPS of $BaFe_2As_2$[26] (a) and calculated XPS VB of $CaFe_2As_2$ (b).

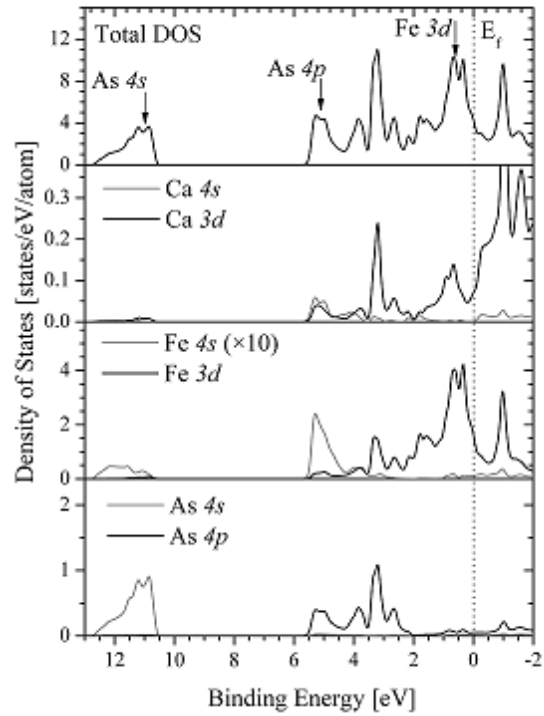

Fig. 1

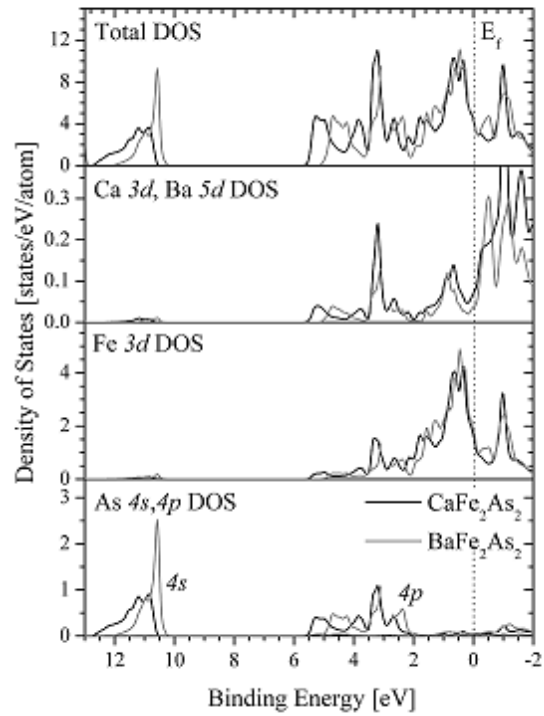

Fig. 2

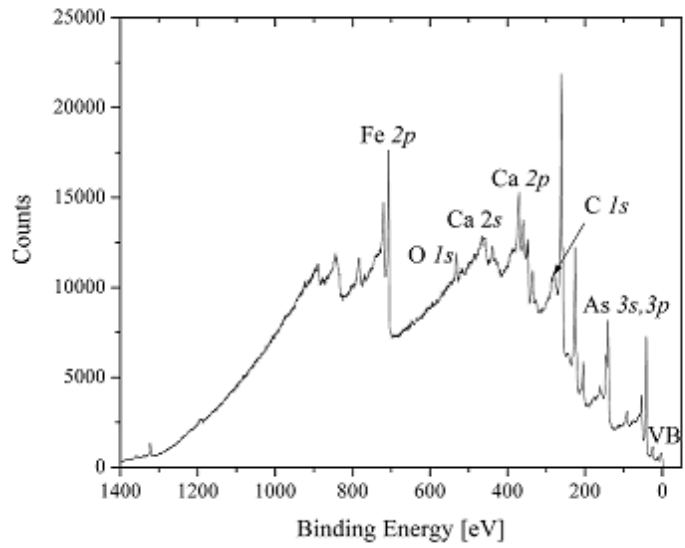

Fig. 3

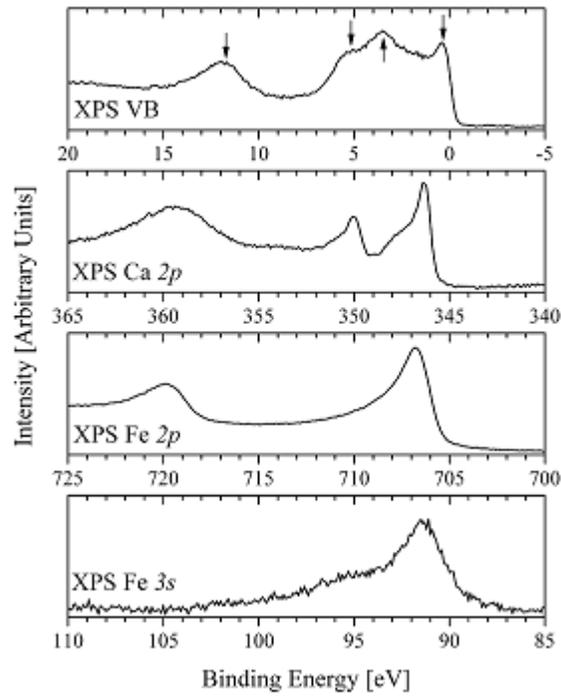

Fig. 4

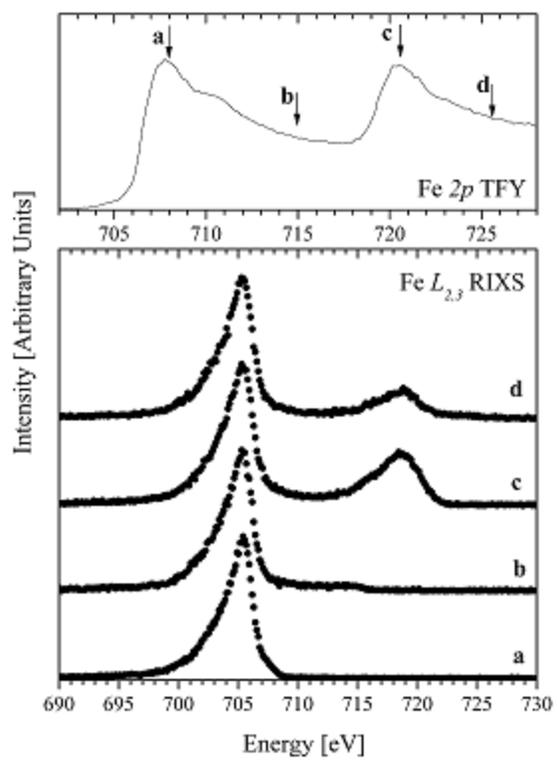

Fig. 5

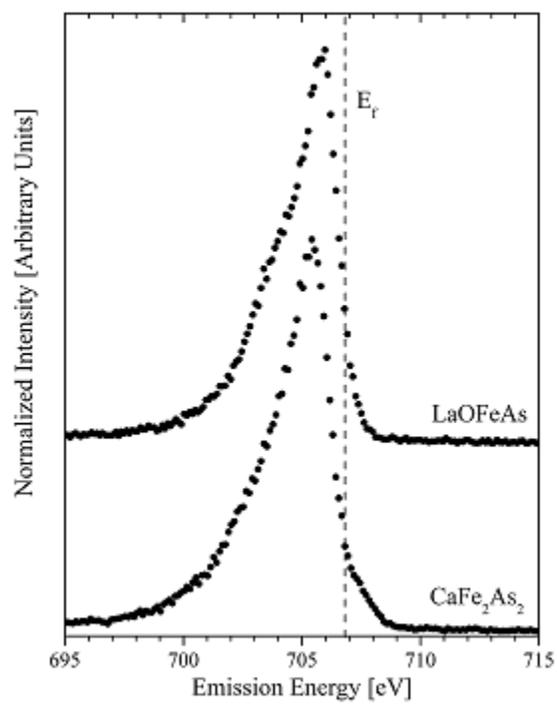

Fig. 6

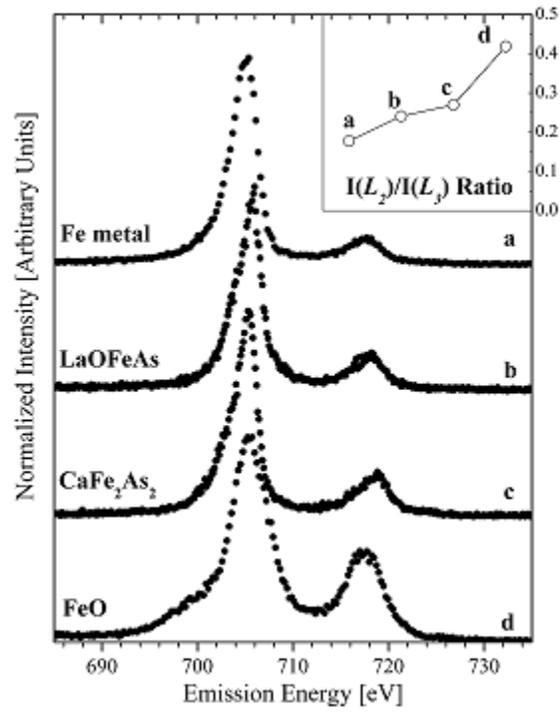

Fig. 7

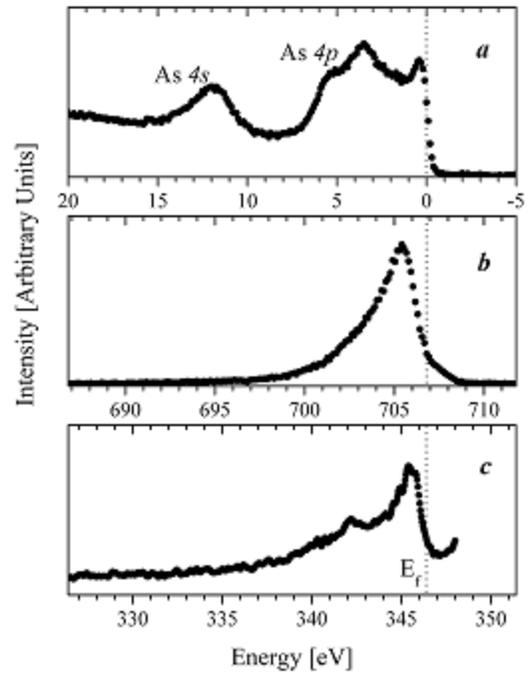

Fig. 8

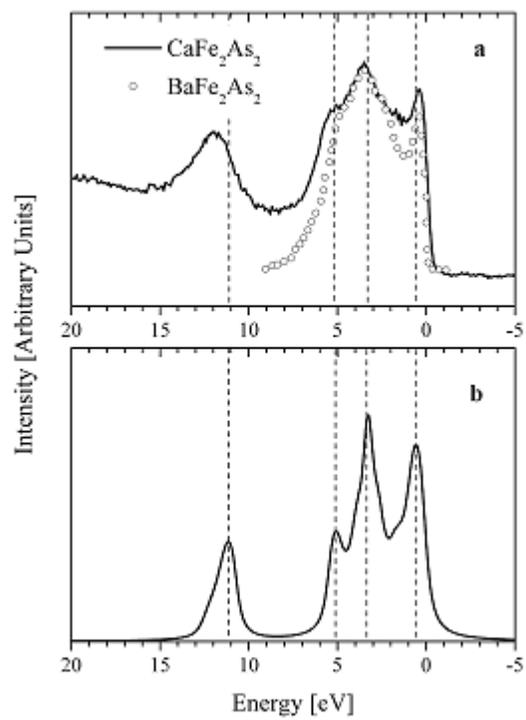

Fig. 9